# Modulation instability of structured-light beams in negative-index metamaterials

Salih Z. Silahli, Wiktor Walasik and Natalia M. Litchinitser

Electrical Engineering Department, University at Buffalo, The State University of New York, Buffalo, NY 14260
E-mail: natashal@buffalo.edu

**Abstract**
One of the most fundamental properties of isotropic negative-index metamaterials, namely opposite directionality of the Poynting vector and the wavevector, enable many novel linear and nonlinear regimes of light-matter interactions. Here, we predict distinct characteristics of azimuthal modulation instability of optical vortices with different topological charges in negative-index metamaterials with Kerr-type and saturable nonlinearity. We derive an analytical expression for the spatial modulation-instability gain for the Kerr-nonlinearity case and show that a specific condition relating the diffraction and the nonlinear lengths must be fulfilled for the azimuthal modulation instability to occur. Finally, we investigate the rotation of the necklace beams due to the transfer of orbital angular momentum of the generating vortex on the movement of solitary necklace beams. We show that the direction of rotation is opposite in positive- and negative-index materials.

## 1. Introduction

Artificially designed metamaterials exhibit unusual properties that cannot be obtained in nature. Among these metamaterials, negative-index metamaterials (NIMs) attract special interest. NIMs are characterized by negative dielectric permittivity $\varepsilon$ and magnetic permeability $\mu$ [1–7] in a certain frequency range and allow for wave propagation. An important property of the wave propagation in NIMs is that the Poynting vector points in the direction opposite of the wave vector. In fact, this property is often considered to be the most general definition of negative-index materials. Using Maxwell's equations for media with $\varepsilon$ and $\mu$, Veselago predicted that the right-handed triplet of vectors $\mathbf{E}$ (electric field), $\mathbf{H}$ (magnetic field), and $\mathbf{k}$ (wavevector) in conventional, positive index material (PIM) changes to the left-handed triplet in NIM, leading to negative refraction of light beams [1].

Antiparallel wave and Poynting vectors in NIMs have been shown to manifest in many remarkable linear and nonlinear optical phenomena [8–23], including negative refraction, amplification of the evanescent waves, new "backward" phase-matching conditions for nonlinear optical interactions [8,24–32], surface and guided waves regimes unattainable in conventional waveguides, and new types of temporal and spatial solitons [33–40].

To date, a majority of the studies of nonlinear light-matter interactions in NIMs have focused on conventional Gaussian beams or pulses. However, recently it was realized that both linear and nonlinear optics of more complex structured light beams, containing phase or polarization singularities, may offer new opportunities for modern photonics, including multidimensional communication systems, nanoscale imaging, and quantum information processing [41–48].

Here, we show that the well-known nonlinear phenomenon of modulation instability (MI) reveals even more remarkable characteristics when structured light beams (optical vortices) with different topological charges propagate in NIMs with Kerr or saturable nonlinearity. A modulation instability phenomenon reveals itself as an exponential growth of weak perturbations when an intense pump beam propagates inside a nonlinear medium [49]. While some of the first studies of this effect date back to the 1960s, this field continues to develop rapidly these days. Over the years, self-focusing and related phenomena of beam filamentation, soliton propagation, and necklace beam formation were studied in detail and in some cases were demonstrated experimentally in various nonlinear media [50–78].

In this paper, we study the propagation of structured light in a nonlinear negative-index medium. We derive an analytical expression for the spatial MI gain for the Kerr-nonlinearity case and show that a specific condition relating the diffraction and the nonlinear lengths must be fulfilled for the azimuthal MI to occur. We confirm the analytical results by numerical solution of the nonlinear Schrödinger equation and estimate the loss level necessary to observe the MI of optical vortices in NIMs at physically attainable power levels. Moreover, we show that the necklace beams originating from the same vortex propagating in positive- and negative-index materials rotate in the opposite direction.

## 2. Linear Stability Analysis

The nonlinear Schrödinger equation governing the evolution of the slowly varying electric-field envelope $A$ in NIMs is derived in Ref. [74]. For the continuous wave propagation studied here, we neglect the dispersion and self-steeping terms Eq. (10) in Ref. [74] and obtain the following nonlinear Schrödinger equation:

$$\frac{\partial A}{\partial z} = \left[\frac{i}{2k}\nabla_\perp^2 + kn_i + if(|A|^2)\right]A, \quad (1)$$

where $k = n_r \omega_0/c$ is the wavenumber, $\omega_0$ denotes the light angular frequency, $c$ is the speed of light in vacuum, and $\nabla_\perp^2$ denotes the transverse Laplacian. The refractive index of the NIM is given by $n = n_r + in_i$, where both $n_r$ and $n_i$ are real. The function $f(|A|^2)$ describes the nonlinear response of the material. In the linear stability analysis, we consider a Kerr-type nonlinear response in the form $f(|A|^2) = C_{\text{nl}}|A|^2$, where the nonlinear coefficient $C_{\text{nl}}$ is defined as

$$C_{\text{nl}} = \omega_0 \text{Re}[\mu_r]\chi^{(3)}/2cn_r. \quad (2)$$

Here, $\mu_r$ stands for the relative magnetic permeability of the medium, and $\chi^{(3)}$ is the third-order nonlinear susceptibility.

Due to the rotational symmetry of the beam, it is convenient to rewrite Eq. (1) in a cylindrical coordinate system

$$i\frac{\partial A}{\partial z} = \frac{i}{2k}\left(\frac{1}{r}\frac{\partial A}{\partial r} + \frac{\partial^2 A}{\partial r^2} + \frac{1}{r^2}\frac{\partial^2 A}{\partial \theta^2}\right) + iC_{\text{nl}}|A|^2 A. \quad (3)$$

For the purpose of linear stability analysis, we have neglected the loss term.

We start with performing a linear stability analysis of the vortex beam propagating in a Kerr medium. First, we analytically find steady-state solutions of Eq. (1) using the methodology described in detail in Ref. [56]. Then, we apply azimuthal perturbation to one of the steady-state solutions found following the lines described in Refs. [56,78]. The perturbations are applied only to the field distribution in the azimuthal direction, which allows us to omit the two first terms in the transverse Laplacian in Eq. (3). The perturbations are applied only to the azimuthal field distribution taken at the radial distance for which the intensity is constant $A_o(\theta) = A(r = r_m, \theta)$, where $r_m$ defines the mean radius of the steady-state solution and it is calculated using Eq. (21) in Ref. [56]. The perturbed field distribution is given by

$$A(z,\theta) = \left(|A_o| + a_1 e^{-i(M\theta+\mu z)} + a_2^* e^{i(M\theta+\mu^* z)}\right)e^{i\lambda z+im\theta}, \quad (4)$$

where $|A_o|$ is the amplitude of the steady-state solution; $a_1, a_2$ are the amplitudes of small perturbations ($a_1, a_2 \ll |A_o|$), and $m$ and $M$ are azimuthal indices of the steady-state solution [topological vortex charge ($m = 0, 1, 2, ...$)] and the perturbation, respectively. $\lambda$ is the nonlinear propagation constant of the steady-state solution, and $\mu$ is the propagation constant correction for the perturbation.

Substituting Eq. (4) into Eq. (3), and linearizing the equation in perturbation amplitudes, yields an eigenvalue problem for $a_1$ and $a_2$:

$$a_1(\lambda - \mu) + \frac{(m-M)^2}{2kr_m^2}a_1 - C_{\text{nl}}A_0^2(2a_1 + a_2) = 0, \quad (5)$$

$$a_2(\lambda + \mu) + \frac{(m+M)^2}{2kr_m^2}a_2 - C_{\text{nl}}A_0^2(a_1 + 2a_2) = 0. \quad (6)$$

Solving Eqs. (5)–(6) we obtain two expressions: (i) for the nonlinear propagation constant:

$$\lambda = -\frac{m^2}{2kr_m^2} + C_{\text{nl}}|A_0|^2 \quad (7)$$

and (ii) for the propagation constant correction $\mu$ related with the MI

$$\mu = -\frac{mM}{kr_m^2} \pm \sqrt{\left(\frac{m^2+M^2}{2kr_m^2} + \lambda - 2b\right)^2 - b^2}, \quad (8)$$

where we introduce the nonlinear parameter $b = C_{\text{nl}}A_0^2$. The MI growth rate is characterized by the imaginary part of $\mu$:

$$Im(\mu) = Im\sqrt{\frac{M^4}{4k^2 r_m^4} - \frac{M^2}{kr_m^2}b} \quad (9)$$

where the simplifications were made using Eq. (7). Equation (10) allows us to predict the MI gain as a function of the modulation azimuthal index $M$, the vortex mean radius $r_m$ (which is related with the vortex charge $m$ and the nonlinear propagation constant $\lambda$), and the nonlinear parameter $b$.

While Eq. (10) gives the absolute magnitude of the MI gain, it is instructive to derive the equation for the normalized gain $g$. By introducing a normalized azimuthal perturbation index $\kappa = M/(2\sqrt{|kb|})$ in Eq. (10) and dividing both its sides by $|b|$ we obtain:

$$g = Im\sqrt{\frac{4\kappa^4}{r_m^4} - sgn(n_r)\frac{4\kappa^2}{r_m^2}sgn(\chi^{(3)})}. \quad (11)$$

The condition sufficient for the existence of the MI is $r_m^2/\kappa^2\, sgn(n)sgn(\chi^{(3)}) > 1$. From this condition, we draw two conclusions. (i) The MI is possible when the sign of the real part of the refractive index $n_r$ is the same as the sign of the third-order susceptibility $\chi^{(3)}$. This implies that for NIM the MI should occur for $\chi^{(3)} < 0$. (ii) Using the definition of $\kappa$ and $b$ the condition $r_m^2/\kappa^2 > 1$ can be rewritten in terms of the diffraction length $L_D = \pi |n_r| (2r_m)^2 / \lambda_0$ and the nonlinear length [49] $L_{NL} = (C_{nl}|A_0|^2)^{-1}$ as

$$M^2 < \frac{L_D}{L_{NL}} \quad (12)$$

For given parameters of the input beam, Eq. (12) allows one to identify the maximum value of the perturbation index $M$ for which the modulation instability gain is nonzero.

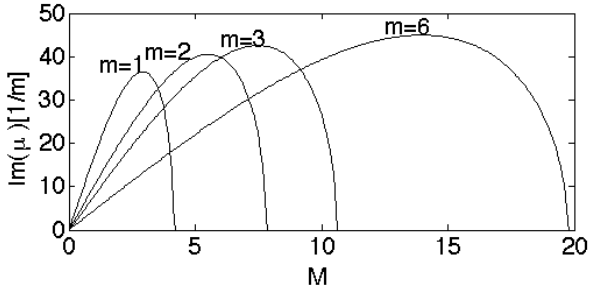

**Figure 1.** Azimuthal modulation instability gain Im($\mu$) as a function of the perturbation azimuthal index $M$, for different topological charges $m$ of the initial steady-state vortex solution for the Kerr nonlinearity case.

Equation (10) is used in the following to predict the MI gain for vortices propagating in the NIM medium. We study the light with the wavelength $\lambda_o = 2\pi c/\omega_0 = 1.405$ μm, where the figure of merit $|n_r/n_i|$ for the double-fishnet negative-index photonic metamaterial proposed in Ref. [79] is the highest and the refractive index is $n = -1 + 0.3i$. The corresponding permeability is $\mu_r = -0.64 + 0.42i$. Here, we consider the NIM with Kerr nonlinearity that might be obtained, for instance, by using a highly nonlinear dielectric, such as chalcogenide glass, for the dielectric layers of the NIM. The third-order susceptibility used here $\chi^{(3)} = -3.61 \cdot 10^{-18}$ [m$^2$/V$^2$] is typical for chalcogenide glasses [80]. Figure 1 shows the gain curves Im ($\mu$) as a function of the perturbation azimuthal index $M$ for different values of the vortex charge $m$. For each of the vortex topological charges, the total power of the steady-state solution is estimated analytically:

$$P_m = \frac{2^{2m-1}m!(m+1)!}{(2m)!Re(C_{nl})}\varepsilon_0 c\lambda_0, \quad (13)$$

where $\varepsilon_0$ is the vacuum permittivity. For this power, there exists a family of steady-state solutions characterized by the nonlinear propagation constant $\lambda$. Different values of $\lambda$ correspond to different widths of the beam $r_m$ and different peak amplitudes $|A_o|$. In the following, we choose $\lambda = 17$ [1/m], which results in typical beam widths between 200 and 600 μm.

We observe that the MI gain is higher as vortex charge $m$ increases. As a result, for a fixed value of the nonlinear propagation constant $\lambda$, the MI onsets at a shorter propagation distance as the vortex charges increases. Additionally, we observe that the number of peaks associated with the perturbation azimuthal index $M$ for which Im ($\mu$) is maximal, is larger for the higher topological charges $m$.

### 3. Numerical propagation

The theoretical predictions based on the analytical expression for the MI gain [Eq. (10)] are verified by a direct numerical solution of a three-dimensional nonlinear Schrödinger equation [Eq. (1)]. This equation is solved using the split-step Fourier algorithm [82,83] in order to study the vortex dynamics. Values of losses in the metamaterials designed up to date are prohibitively large and do not allow for the propagation long enough to observe the MI in optical vortices with values of peak intensity below the damage threshold of the typical optical materials. Therefore, in the numerical simulations we neglect losses in order to be able, at least qualitatively, to predict the MI of vortices in NIMs. In the discussion of the results we estimate the level of loss allowing for an experimental observation of MI.

The input field in the numerical algorithm

corresponds to the steady-state solution of Eq. (1):

$$A(r, \theta, z = 0) = A_m (r/\omega_m)^m \exp^{-r^2/(2r_m^2) + im\theta}, \quad (14)$$

where the width $\omega_m$ for the $m$th order stable vortex is related to the mean radius as $\omega_m = r_m/(m+1)$ and the amplitude $A_m$ can be deduced from the total power of the stable solution given by Eq. (13). In order to accelerate the growth of the MI, 10% of random noise is added to the input field.

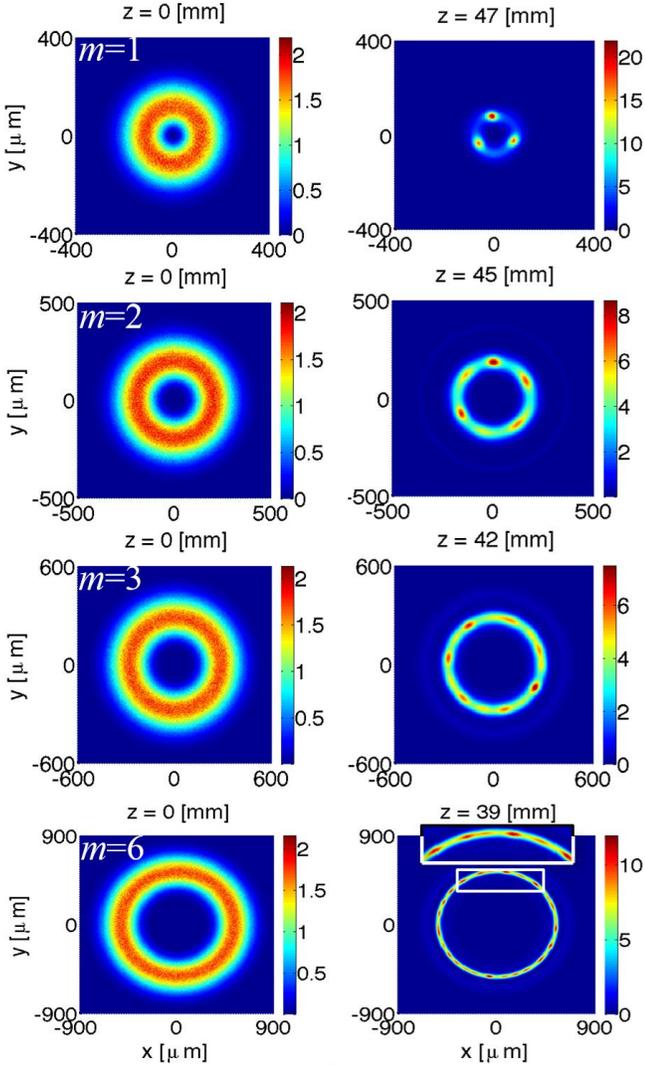

**Figure 2.** Distribution of the electric field intensity $|A^2|[10^{13} \text{ V}^2/\text{m}^2]$ presenting the dynamics of the necklace beams for different topological charges $m = 1, 2, 3$ and 6 in the NIM for Kerr nonlinearity. Zoom of the region marked in white is presented in the inset.

Figure 2 shows the dynamics of the light propagation of vortex beams with various topological charges $m$ in the NIM with Kerr-type nonlinearity. The first column presents the steady-state solution with the noise, and the second column presents the generated necklace beams immediately after the onset of the MI breakup. There is good agreement between the analytical predictions shown in Fig. 1 and the numerical simulations in Fig. 2 both in terms of the number of necklace beams generated and the distance at which the MI onsets. This distance, inversely proportional to the MI gain $\text{Im}(\mu)$, decreases with the increase of the vortex topological charge, as predicted analytically in Fig. 1.

The values of the power corresponding to the steady-state solutions in the NIM system calculated using Eq. (12) are $P_1 = 1.44 \text{ kW}$, $P_2 = 2.88 \text{ kW}$, $P_3 = 4.58 \text{ kW}$, and $P_6 = 11.2 \text{ kW}$. These values were obtained neglecting losses. For these values, we indeed observe the stable propagation of the vortex beam over the initial distance of around 150 mm. After that, the MI onsets and the necklace beams are generated. However, this propagation distance requires a long NIM sample. In order to shorten the necklace beam generation distance, we use the input beams with powers higher than those of the steady-state solutions. For each of the charges, the power is increased by the same factor of 2.4 while keeping all the other parameters unchanged. For this reason in the numerical simulations, the width of the beams oscillates during the propagation. The input power levels in the numerical simulations are $P_1 = 3.46 \text{ kW}$, $P_2 = 6.93 \text{ kW}$, $P_3 = 11 \text{ kW}$, and $P_6 = 26.88 \text{ kW}$. The results of the simulations using these power levels are shown in Figs. 2–5.

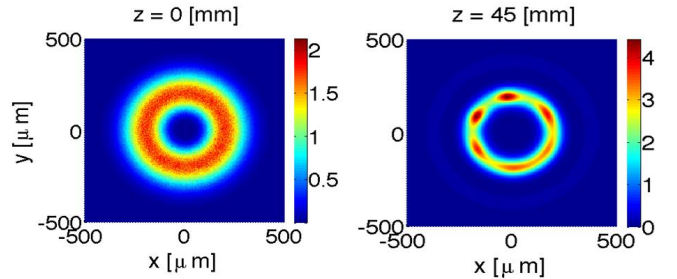

**Figure 3.** Distribution of the electric field intensity $|A^2|[10^{13} \text{ V}^2/\text{m}^2]$ presenting the dynamics of the necklace beam with the topological charge $m = 2$ in the NIM for Kerr nonlinearity. The loss associated with the imaginary part of the refractive index $n_i = 5 \cdot 10^{-7}$ is taken into account.

All the previous results were obtained neglecting losses of the NIM. Here, we want to estimate the maximum value of loss that would not prohibit the MI induced necklace beam generation. We assume that the MI will not be suppressed if the light intensity decreases $e$ times over the distance required for MI onset in the absence of losses. For the charge 2 vortex studied here, this theoretical estimation results in the imaginary part of the refractive index $n_i = 5 \cdot 10^{-7}$. The numerical simulations presented in Fig. 3 confirm this analytical estimation. For $n_i = 5 \cdot 10^{-7}$, the MI onsets at the same distance as in the lossless case, but the resulting peak intensity is two times lower than in the lossless case. Numerical propagation of the same input beam with different values of $n_i$ shows that for higher losses, the MI onset occurs for longer propagation distances or does not happen at all if the losses are too large. The value of $n_i$ for which the MI still can be observed is six orders of magnitude lower than the values attainable in to-date metamaterials. This shows great progress that needs to be made in the field of metamaterials in order to enable the experimental observation of necklace beam generation in NIMs.

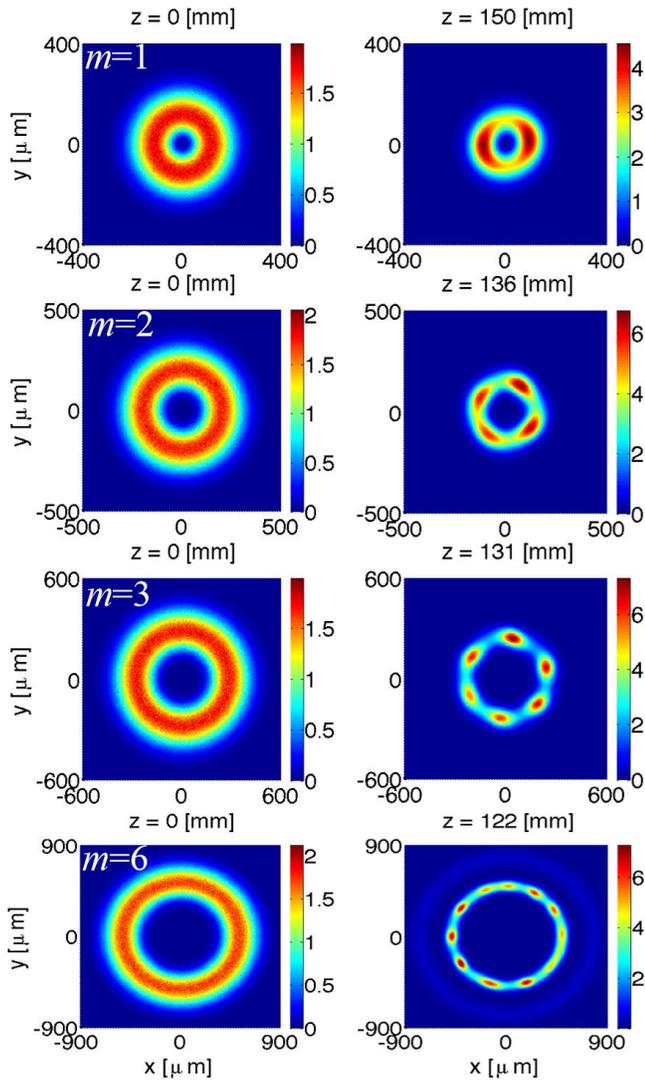

**Figure 4.** Distribution of the electric field intensity $|A^2|[10^{13}\,V^2/m^2]$ presenting the dynamics of the necklace beams for different topological charges $m = 1, 2, 3,$ and $6$ in the NIM for saturable nonlinearity.

The orbital angular momentum of the vortex beam is partially transferred onto the rotation of the necklace beam pattern [81]. For positive vortex charges in the PIM, the beams rotate counterclockwise, while propagating towards the observer. However, the direction of rotation is reversed by either changing the vortex charge or by propagating a vortex with the same charge in the NIM. In the latter case, the energy flow direction is reversed due to the fact that the Poynting vector and the wavevector are antiparallel. Here, we want to demonstrate this effect by propagating the same input vortex with PIM and NIM with opposite signs of $n_r$ and, consequently, Re($\mu_r$). This effect is difficult to observe in materials with Kerr-type nonlinearity, because of the rapid collapse of the generated beams after the IM onset. Therefore, we study the beam generation in materials with saturable nonlinearity given by

$$f(|A|^2) = \frac{C_{nl}|A|^2}{1 + |A|^2/A_{sat}^2} \quad (15)$$

where the saturation amplitude is chosen to be $A_{sat}^2 = 5 \cdot 10^{13}\,[V^2/m^2]$. The results of the propagation of vortices in NIM with saturable nonlinearity are presented in Fig. 4. All the other parameters and the input beam profiles are the same as those used in Fig. 2. The comparison of Figs. 2 and 4 shows that there is no qualitative difference between the MI onset and necklace beam generation in the case of Kerr and saturable nonlinearities. However, there are qualitative differences: the number of necklace beams generated is lower and the distance required for the MI onset is longer in the case of the saturable nonlinearity. Moreover, due to a weaker self-focusing effect, the diameter of a solitary necklace beam is larger for the saturable nonlinear medium.

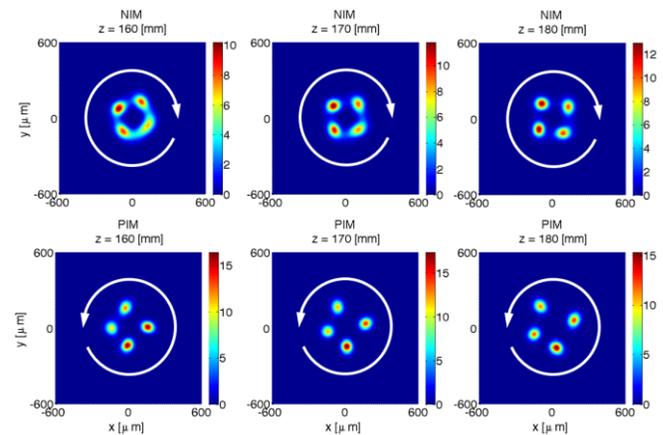

**Figure 5.** Distribution of the electric field intensity $|A^2|[10^{13}\,V^2/m^2]$ presenting the rotation of the necklace beams with topological charge $m = 2$ during the propagation in the NIM (top row) and PIM (bottom row) for saturable nonlinearity. White arrows indicate the direction of rotation.

Figure 5 presents the comparison of the propagation of a charge 2 vortex in PIM and NIM. We can clearly see that the direction of rotation in the two cases is opposite. The video presenting the beam evolution during the propagation can be found in the Supplementary materials. Due to the use of saturable nonlinearity, the beams do not collapse after MI onset, and they propagate as stable solitary beams.

## 4. Conclusion

In summary, we investigated the azimuthal modulation instability of vortex beams carrying optical angular momentum in negative-index metamaterial and predict the formation of necklace beams. We have shown both analytically and numerically that, if the loss is sufficiently low, necklace beams with various numbers of beams can be generated using different charges of the input vortex beams. The analytical condition for the modulation instability to occur has been derived in terms of signs of the real parts of the refractive index and magnetic permeability, and in terms of diffraction and nonlinear lengths. Moreover, we have shown numerically that the generation of necklace beams in negative-index material with saturable nonlinearity is qualitatively similar to the case of Kerr-type nonlinearity. For saturable nonlinearity, the necklace beams do not collapse, and we were able to observe their rotation during the propagation. We have demonstrated that the rotation direction is opposite in the positive- and negative-index materials for the same input vortex parameters, as expected.

## Acknowledgements

This work was supported by Army Research Office (ARO) grants W911NF-11-1-0297 and W911NF-15-1-0146).